\documentclass[aps,prl,showpacs,groupedaddress,twocolumn,preprintnumbers,amsmath,amssymb]{revtex4}

\usepackage{graphicx}% Include figure files

\begin{document}

\title{Generation of Dicke States with Phonon-Mediated Multi-level Stimulated Raman Adiabatic Passage}% Force line breaks with \\

\author{Atsushi~Noguchi}
\email[]{noguchi@qe.ee.es.osaka-u.ac.jp}
\author{Kenji~Toyoda}
\author{Shinji~Urabe}
\affiliation{%
Graduate School of Engineering Science, Osaka University, 1-3 Machikaneyama, Toyonaka, Osaka, Japan}

\date{\today}% It is always \today, today,
             
\begin{abstract}
We generate half-excited symmetric Dicke states of two and four ions.
%($\mid\!\mathrm{D}^1_2\rangle$ and $\mid\!\mathrm{D}^2_4\rangle$)
We use multi-level stimulated Raman adiabatic passage (STIRAP) whose intermediate states are phonon Fock states.
This process corresponds to the spin squeezing operation and half-excited Dicke states are generated during multi-level STIRAP.
This method does not require local access for each ion or the preparation of phonon Fock states. Furthermore, it is robust since it is an adiabatic process.
We evaluate the Dicke state using a witness operator and determine the upper and lower bounds of the fidelity without using full quantum tomography.
\end{abstract}

\maketitle

Genuine multipartite entanglement has been extensively investigated in quantum information science \cite{1}.
It is very difficult to generate genuine multipartite entanglement because the entangled states of many qubits decay with the square of the qubit number \cite{1}.
To date, the Greenberger--Horne--Zeilinger (GHZ) state of 14 qubits has been generated \cite{1} using a M$\o$lmer-S$\o$rensen gate \cite{2}; however, the half-excited symmetric Dicke states that form another important group of entangled states have been generated by photonic qubits with only four and six photons \cite{3,4}.
No robust and efficient method has been demonstrated for generating Dicke states using more than two trapped ions\cite{5,a}.

Dicke states are generally the simultaneous eigenstates of the total angular momentum operator and its projection along one direction.
One subgroup of Dicke states is symmetric under permutation of qubits:
\begin{equation*}
\mid\! D_{n(z)}^{m} \rangle =\frac{1}{\sqrt{{}_nC_m}}\sum _k \hat{P}_k\mid \uparrow _z\uparrow _z\dots \uparrow _z\downarrow _z\downarrow _z\dots\downarrow _z\rangle
\end{equation*}
where $P_k$ is the permutation operator and ${}_nC_m =n!/[m! (N-m)!]$.
In this letter, we refer to these states as symmetric Dicke states.
Here, we focus on the half-excited symmetric Dicke states ($\mid \!\mathrm{D}^n_{2n(z)}\rangle$), which are difficult to generate by addressing local qubits with an array of laser pulses \cite{6} or post-selection by photodetection with atomic ensemble \cite{7}.
These multi-partite entangled states are not only important for quantum information science, they are also useful for precise measurements such as Heisenberg-limited measurements \cite{8,9,10,11} because these states are related to the squeezed vacuum state in quantum optics through the two-excitation correlation \cite{12}.
Moreover, half-excited symmetric Dicke states can form decoherence-free entangled states with a dressing field \cite{15}.

In this letter, we use phonon-meditated multi-level stimulated Raman adiabatic passage (STIRAP) \cite{b,c} to generate half-excited symmetric Dicke states of two and four ions.
%The adiabaticity and simpleness of this method make the fidelity of generated states high.
We evaluate the four-ion Dicke state with a witness operator and the fidelity without using full quantum tomography.

%%%% Figure 1%%%%
\begin{figure}[b]
   \includegraphics[width=5cm,angle=0]{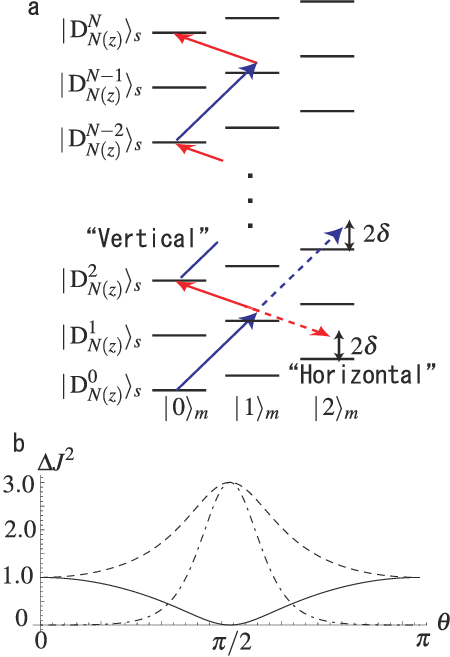}
\caption{(a) Energy-level diagram of phonon-mediated multi-level STIRAP. Vertical ladders indicate spin states and horizontal arrays represent phonon Fock states.
(b) Evolution of spin noise for four ions. Solid, dashed, and dot-dashed curves indicate $\Delta J_x^2$, $\Delta J_y^2$, and $\Delta J_z^2$, respectively. %$\theta$ is the parameter of the dark states.
}
\label{fig1}
\end{figure}
%%%%%%%%%%%%%

We consider the case where ions are cooled to the ground state and are irradiated by two color laser beams that are detuned from the blue and red sideband transitions for the M$\o$lmer S$\o$rensen interaction \cite{2}.
In the interaction picture, the Hamiltonian is expressed by
\begin{equation*}
\hat{H}=\frac{\eta \Omega _r}{2} (\hat{a} \hat{J}_+ e^{-i\delta t}+\hat{a}^\dagger \hat{J}_-e^{i\delta t})+\frac{\eta \Omega _b}{2}(\hat{a}^\dagger \hat{J}_+e^{i\delta t}+\hat{a}\hat{J}_-e^{-i\delta t})
\end{equation*}
where $\hat{J}_{\pm}$ ($=\sum_i \hat{\sigma} _\pm^{(i)}$, where $\hat{\sigma} _\pm^{(i)}$ is the Pauli operator for the $i$th ion) is the global spin flip operator and $\hat{a} (\hat{a}^\dagger)$ is the annihilation (creation) operator for the center of motion of the ion chain, and $\delta$ is the detuning of the blue and red sideband transitions.
We begin with the initial state $\mid \downarrow\downarrow \dots \downarrow\rangle _s$ $\mid 0\rangle _p$, where $\mid\! 0\rangle _p$ is the phonon vacuum state.
This initial state is obviously the dark state of this Hamiltonian when $\Omega _b=0$ and we adiabatically change the amplitudes of the two color beams to transfer the population to the other dark states ($\mid \uparrow\uparrow \dots \uparrow\rangle _s \mid 0\rangle _p$, where $\Omega _r =0$).
Here, we calculate the dark states and show that the half-excited symmetric Dicke states are the dark states of the Hamiltonian.

We represent the Hamiltonian by the eigenstates of the global spin operator $\hat{\bm{J}}^2$ and $\hat{J}_z$.
Due to the symmetry of the Hamiltonian, 
we can consider the symmetric subspace of the whole $N$ qubit Hilbert space, whose basis is 
$\mid\!\mathrm{D}^{m}_{N(z)} \rangle _s\mid\! n\rangle _p$, where $N$ is the number of ions and $\mid\!\! n\rangle _p$ is the phonon Fock states.
Fig. 1(a) shows the interaction for the lower phonon states ($\mid \!0\rangle _p, \mid\! 1\rangle_p$ and $\mid\! 2\rangle _p$).
We assume that the red and blue sideband transitions have the same magnitudes for one-photon detuning $\delta$ and that the ``vertical" Raman transitions [Fig. 1(a)] are on resonance relative to two-photon detuning.
On the other hand, there are other Raman transitions, which are ``horizontal" Raman transitions [Fig. 1(a)]; these transitions change the mean phonon number.
For the case in which vertical Raman transitions are on resonance, horizontal Raman transitions have two-photon detuning as $2\delta$.
When $2\delta \gg\eta\Omega _r,\eta\Omega _b$, the horizontal Raman transitions can be ignored and the interaction can be expressed in Fig. 1(a) corresponding to the multi-level Raman transitions \cite{a}.
The Raman couplings are not independent and they have different strengths due to the cooperation effects of symmetric Dicke states.
The coupling strengths of the cooperation effects are expressed by
\begin{equation*}
R_{m}\equiv _s\langle \mathrm{D}^{m+1}_{N(z)}\mid\hat{J}_+\mid\!\mathrm{D}^m_{N(z)}\rangle _s =(N-m)\sqrt{\frac{_N\mathrm{C}_{m}}{_N\mathrm{C}_{m+1}}},
\end{equation*}
and the Hamiltonian is expressed as follows in rotation frames
\begin{equation*}
H=
\begin{pmatrix}
0&R_0\eta\Omega _b&0&&\cdots &0\\
R_0\eta\Omega _b^*&\delta &R_1\eta\Omega _r& &&0\\
0&R_1\eta\Omega _r^*&0 & &&0\\
%0&0&R_2\eta\Omega _b^*& &&0\\
\vdots & & &\ddots &&\vdots \\
0& &          & &\delta &R_{N-1}\eta \Omega _r\\
0& &\cdots & &R_{N-1} \eta\Omega_r^* &0
\end{pmatrix}
\end{equation*}
with a basis of
$\{\mid\!\!\mathrm{D}_{N(z)}^0\rangle _s\mid\!\! 0\rangle_p ,\mid\!\!\mathrm{D}_{N(z)}^1\rangle _s\mid\!\! 1\rangle_p , \dots ,$ $\mid\!\mathrm{D}_{N(z)}^{N-1}\rangle _s\mid\! 1\rangle_p , \mid\!\mathrm{D}_{N(z)}^N\rangle _s\mid\! 0\rangle_p\}$.

In this letter, we consider only even numbers of ions.
This allows us to calculate the dark state $\mid\! \psi _d (\Omega _r,\Omega _b)\rangle$ 
%by making all transition amplitudes disappear with each other
\cite{b,c} as
\begin{equation*}
\mid\! \psi _d(\Omega _r,\Omega _b)\rangle =A\sum _{i=0}^{N/2} C_{i}\Omega _b^{i}\Omega _r^{N/2-i} \mid\!\mathrm{D}^{2i}_{N(z)}\rangle _s\mid\! 0\rangle _m,
\end{equation*}
\begin{eqnarray*}
C_{0}&=&1\\
C_{i}&=& (-1)^i\prod _{j=1} ^{i}\frac{R_{2j-2}}{R_{2j-1}},\text{@}(i=1,\dots \frac{N}{2})
\end{eqnarray*}
where $A$ is a normalizing constant.
The dark state exists only for the motional ground state.
%Therefore, the ions initially must be cooled to the motional ground state.

%When we change the parameters $\Omega _r\rightarrow\Omega _r-dr$, $\Omega _b\rightarrow\Omega _b+db$,
Since each dark states (except for the initial and final states) is a superposition of states which are connected due to the quadratic raising and lowering operators of $\hat{J}_+\hat{J}_+$ and $\hat{J}_-\hat{J}_-$,
the dark states correspond to the vacuum squeezed state in quantum optics \cite{12,d}.
%\footnote{In the quantum optics, $\hat{H}_{param}=i (\hat{a}^\dagger\hat{a}^\dagger-\hat{a}\hat{a})$ is the spin squeezing Hamiltonian, where $\hat{a} (\hat{a}^\dagger)$ is the annihilation (creation) operator of photon} and are characterized by the variance of global spin operators.
To see these characteristics, we calculate the spin noise for the four-ion case.
Fig. 1(b) shows the dependence of the variance of the global spin operator ($\Delta\hat{J}_x,\Delta\hat{J}_y,\Delta\hat{J}_z$) for dark states on the parameter $\theta$ ($\Omega _b=1-\cos\theta, \Omega _r=1+\cos\theta$).
For $0<\theta <\frac{\pi}{2}$ or $\frac{\pi}{2}<\theta <\pi$,
the spin noises are squeezed or anti-squeezed and when $\theta =\frac{\pi}{2} (\Omega _b=\Omega _r)$, spin noise in one direction goes to zero and the half-excited symmetric Dicke state in the $x$ direction ($\mid\!\mathrm{D}^2_{4(x)}\rangle =\hat{R}_y(\pi /2)\mid\!\mathrm{D}^2_{4(z)}\rangle $) is generated, where $\hat{R}_y(\theta)$ is the global rotation operator about the $y$ axis.
Even for the general case of $2n$ ions, the half-excited symmetric Dicke state ($\mid\!\mathrm{D}^n_{2n(x)}\rangle$) is generated when $\Omega _b=\Omega _r$
because the dark state $\mid\!\psi _d(\Omega ,\Omega)\rangle$ is the eigenstate of the angular momentum operator in the $x$ direction ($\hat{J}_x=(\hat{J}_++\hat{J}_-) /2$)
\begin{equation*}
\hat{J}_x \mid\!\psi _d(\Omega ,\Omega)\rangle =0.
\end{equation*}
This population transfer using the dark states of phonon-mediated multi-level Raman transitions is robust against certain fluctuations because the quantum state is always in the dark state of the Hamiltonian, as in the STIRAP method \cite{13,14}.
%Moreover the generated states are the decoherence-free entangled state using dressing field\cite{14,15}.

%%%% Figure 2%%%%
\begin{figure}[b]
   \includegraphics[width=5cm,angle=0]{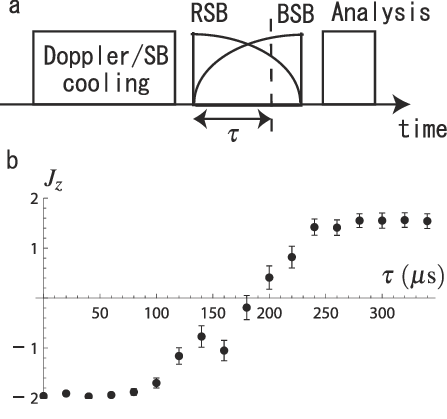}
\caption{(a) Pulse sequences for phonon-mediated STIRAP. We measure the evolution of spin and spin noise by truncating the STIRAP pulse after time $\tau$.
(b) Time evolution of population transfer from the $\mid \downarrow\downarrow\downarrow\downarrow\rangle$ state to the $\mid \uparrow\uparrow\uparrow\uparrow\rangle$ state.}
\label{fig2}
\end{figure}
%%%%%%%%%%%%%

%%%% Figure 3%%%%
\begin{figure}[b]
   \includegraphics[width=5cm,angle=0]{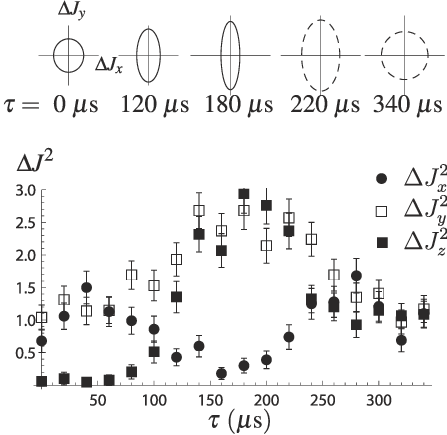}
\caption{
(upper) Visualized spin noise in $XY$ plane, which evolves with phonon-mediated STIRAP. The solid (dashed) curves shows $J_z<0$ ($J_z>0$).
(lower) Time evolution of spin noise. Circles, white squares, and black squares indicate $\Delta J_x^2$, $\Delta J_y^2$ and $\Delta J_z^2$, respectively. $J_x$ ($J_y$ and $ J_z$) is (anti-)squeezed.\\
}
\label{fig3}
\end{figure}
%%%%%%%%%%%%%

We demonstrate phonon-mediated STIRAP with two and four calcium ions trapped in a linear Paul trap.
% whose normal mode frequencies for the center of mass motions are $\{\omega _x,\omega _y, \omega _z\}/2\pi =\{3.1, 3.0, 0.9\} \mathrm{MHz}$. 
Details of the experimental system used are described in our previous studies \cite{15,16}. 
The center-of-mass mode of the axial direction is cooled by sideband cooling to the ground state ($n_{COM}\sim 0.02$), while the other motional modes are cooled only by Doppler cooling.
The qubit is composed of a ground state $\mathrm{S}_{1/2}(m=-1/2)$ and a metastable state $\mathrm{D}_{5/2}(m^\prime =-5/2)$ of a calcium ion and we use $\mathrm{S}_{1/2}-\mathrm{D}_{5/2}$ transitions to excite sideband transitions.
We first initialize all the qubits by optical pumping and irradiate them only with the red sideband laser pulse so that the initial state becomes the dark state.
To operate phonon-mediated STIRAP, we change the amplitude of the sideband pulses, as shown in Fig. 2(a).
The peak Rabi frequency of transitions is $\eta\Omega =2\pi\times 14\mathrm{ kHz}$ and the pulse length is $\tau =340\mathrm{ \mu s}$.
The adiabatic condition is satisfied as $\eta \Omega \tau\sim 5 >1$ \cite{b}.
Fig. 2(b) shows the measured spin projection $\langle\hat{J}_z\rangle$ during population transfer for the four-ion case where a high-efficiency transfer is demonstrated.
To see the characteristics of spin squeeze, the time evolution of the variance of the global spin operator is measured. It is shown in Fig. 3 for four ions.
This shows that spin noises are squeezed or anti-squeezed.

When the two sideband pulses have the same amplitude, the half-excited symmetric Dicke state given by
\begin{eqnarray*}
\mid \mathrm{D}^1_{2(x)}\rangle &=&\frac{1}{\sqrt{2}}(\mid \mathrm{D}^0_{2(z)}\rangle -\mid\mathrm{D}^2_{2(z)}\rangle ),\\
\mid \mathrm{D}^2_{4(x)}\rangle 
 &=&\sqrt{\frac{3}{8}}\mid\mathrm{D}^0_{4(z)}\rangle 
 -\sqrt{\frac{1}{4}}\mid\mathrm{D}^2_{4(z)}\rangle 
 +\sqrt{\frac{3}{8}}\mid\mathrm{D}^4_{4(z)}\rangle
\end{eqnarray*}
is generated.
We evaluate these half-excited symmetric Dicke states for the two- and four-ion cases.

%%%% Figure 4%%%%
\begin{figure}[b]
   \includegraphics[width=5cm,angle=0]{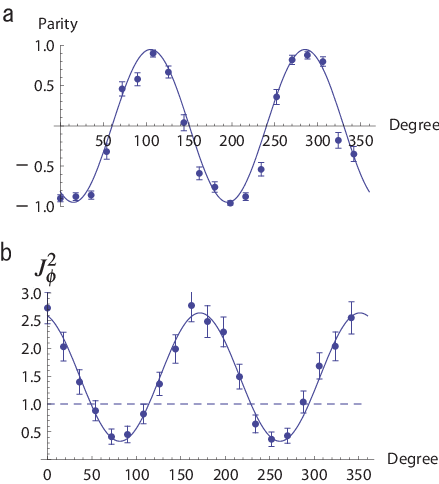}
\caption{(a) Parity oscillation of the Dicke state of two ions.
(b) Squared spin of the half-excited Dicke state of four ions.
%The oscillation of spin noise for Dicke state is observed.
The dashed line indicates shot noise for four ions.
%These two fitting are by the sinusoidal function.
}
\label{fig4}
\end{figure}
%%%%%%%%%%%%%

For the two-ion case, because the Dicke state is the same as one of the Bell states, we can estimate the fidelity only with a global operation and by measuring a parity oscillation.
When $\Omega _r =\Omega _b$, we irradiate the ions with a $\pi /2$ pulse whose phase is $\phi$ and measure the expectation value of a parity operator.
Fig. 4(a) shows the parity oscillation; its amplitude is estimated to be $A_p=0.95 \pm 0.05$ by fitting.
Without the $\pi /2$ pulse, the diagonal elements are measured to be $\{p_{-1},p_0,p_1\}=\{0.516,0.033,0.451\}\pm \{0.011,0.004,0.011\}$.
Using these values, we can obtain the fidelity as $F=\frac{p_{-1}+p_1+A_p}{2}=0.96\pm 0.03$, which is higher than 0.5 \cite{17}.

To evaluate the four-ion Dicke state, we use (i) witness and (2) fidelity methods.
For the four-ion Dicke states, the witness operator is given by \cite{18}
\begin{equation*}
\hat{W}_{ij}=\hat{J}_i^2+\hat{J}_j^2
\end{equation*}
where $i$ and $j$ are any two orthogonal directions.
For the Dicke state in the $x$ direction ($\mid\!\mathrm{D}^n_{2n(x)}\rangle$), we can evaluate its entanglement with the witness $\hat{W}_{yz}$.
After phonon-mediated STIRAP for four ions, we irradiate the ions with a $\pi /2$ pulse whose phase is $\phi$ and measure the expectation values of the square of the spin $J_{\phi}^2$ (Fig. 4(b)).
This maximum peak ($2.64\pm 0.05$) corresponds to $\langle \hat{J}_y^2\rangle$.
On the other hand, $\langle\hat{J}_z^2\rangle$ is measured as $2.82\pm 0.05$ without an analysis pulse.
The witness is then calculated as
\begin{equation*}
\langle\hat{W}_{yz}\rangle =5.46\pm 0.07 >5.23,
\end{equation*}
which is greater than 3-$\sigma$ of the threshold (5.23) of genuine four-partite entanglement \cite{18}.

We also evaluate the Dicke state with the fidelity, which is important to demonstrate that the generated state is closer to the Dicke state than the other genuine four-partite entangled states.
For the Dicke state, the fidelity cannot be exactly determined without full quantum tomography \cite{19}, which requires local access and many measurements \cite{20}.
Here, we evaluate the lower and upper bounds of the fidelity, rather than the fidelity, by using 
only global access and a few measurements.
Details of this method are described in the Appendix.
For the four-ion case, the fidelity($\mathrm{F}$) of the half-excited symmetric Dicke state along the $x$ direction is bounded such that
\begin{equation*}
\frac{\langle \hat{W}_{yz}\rangle}{4}-(\frac{p_{-2}^{(x)}+p_{2}^{(x)}+p_{0}^{(x)}}{2}+\frac{5(p_{-1}^{(x)}+p_{1}^{(x)})}{4})\leq\mathrm{F}\leq p_{0}^{(x)}
\end{equation*}
where $p_{i}^{(x)}$ is the population of the angular momentum state whose spin component along the $x$ direction is $i$.
When $\Omega _r = \Omega _b$, we measure $p_{i}^{(x)}$ as $\{p_{-2}^{(x)},p_{-1}^{(x)},p_{0}^{(x)},p_{1}^{(x)},p_{2}^{(x)}\}=\{0.00,0.03,0.88,0.03,0.03\}\pm \{0.00,0.02,0.03,0.02,0.02\}$.
From these population and witness, we evaluate the fidelity as
\begin{equation*}
0.84\pm 0.03 \leq\mathrm{F}\leq 0.88\pm 0.03
\end{equation*}
Because the maximum fidelity between the GHZ and Dicke states is 3/4 (=$\mid\! \langle \mathrm{GHZ}\!\mid\!\mathrm{Dicke}\rangle\!\mid^2$, where $\mid\!\!\mathrm{GHZ}\rangle =\frac{1}{\sqrt{2}}(\mid \downarrow\downarrow\downarrow\downarrow\rangle +\mid \uparrow\uparrow\uparrow\uparrow\rangle )$, ${\mid \!\mathrm{Dicke}\rangle =\mid \!\mathrm{D}^2_{4(x)}\rangle }$),
this lower bound $(0.84>3/4)$ demonstrates that the generated state is a Dicke state.

In summary, we demonstrate phonon-mediated STIRAP and generate the half-excited Dicke states of two and four ions.
This method requires only the motional ground state and global access to whole ions. Moreover, the adiabatic process, which corresponds to the spin squeezing operation, makes the method robust against certain disturbances and high-fidelity Dicke states can be generated.
The fidelity of these Dicke states are measured as $\mathrm{F}=0.96\pm 0.03$ for two ions and $0.84\pm 0.03\leq\mathrm{F}\leq 0.88\pm 0.03$ for four ions.
Using the witness operator, we show that the generated four-ion state is a genuine four-qubit entangled state and, from the lower bound of the fidelity, we show that this state is the half-excited symmetric Dicke state.
%Phonon-mediated STIRAP can be effectively applied to many ions only if a center of mass mode can be cooled to the ground state.

This work was supported by MEXT Kakenhi ``Quantum Cybernetics" Project and the JSPS through its FIRST Program.
One of the authors (N. A.) was supported in part by the Japan Society for the Promotion of Science.

\section*{Appendix}

We calculate the upper and lower bounds of the fidelity of the half-excited Dicke states ($\mid\!\mathrm{D}^n_{2n}$).
We suppose that the witness ($\langle \hat{W}_{xy}\rangle$) and the population of angular momentum state along the $z$ axis have been measured.
\begin{eqnarray*}
W&=&\langle \hat{W}_{xy}\rangle =\langle \hat{J}_x^2\rangle +\langle \hat{J}_y^2\rangle =\langle \hat{J}^2\rangle -\langle \hat{J}_z^2\rangle\\
P&=&\{ p_{-j_{\mathrm{M}}},p_{-j_{\mathrm{M}}+1},\dots ,p_0,\dots ,p_{j_\mathrm{M}}\}
\end{eqnarray*}
where $j_{M}$ is the maximum projection along the $z$ axis, which is equal to the ion number divided by two.
We consider the diagonal elements of the density matrix based on the angular momentum state along the $z$ axis.
\begin{equation*}
\rho _{j,j_z,\xi}=\langle j,j_z,\xi \mid \hat{\rho}\mid j,j_z,\xi\rangle ,
\end{equation*}
where $\mid\! j,j_z,\xi\rangle$ satisfies $\hat{J}^2\mid\! j,j_z,\xi\rangle =j(j+1)\mid\! j,j_z,\xi\rangle$ and $\hat{J}_z\mid\! j,j_z,\xi\rangle =j_z\mid\! j,j_z,\xi\rangle$.
$\xi$ ($=1,2,\cdots$) is an index that distinguishes between states with the same angular-momentum quantum numbers.
States with $j=j_M$ comprise only these perfectly symmetric states. We then assume that the states 
$\mid \! j =J_M, j_z, \xi =1\rangle$ are symmetric Dicke states.
We can calculate the population P such that $p_{j_z}=\sum _{j,\xi}\rho _{j,j_z,\xi}$.
Note that $\rho _{j_{\mathrm{M}},0,1}$ is the fidelity of the half-excited Dicke state.

For the upper bound, we expand $p_0$ as
\begin{equation*}
p_0=\sum _{j,\xi} \rho _{j,0,\xi} \geq \rho_{j_{\mathrm{M}},0,1} =\mathrm{F}
\end{equation*}
because the diagonal elements of the density matrix are non-negative ($\rho _{j,j_z,\xi}\geq 0$).

For the lower bound, we use the witness operator.
As the witness operator is diagonalized by the angular momentum state along the $z$ axis, we have
\begin{eqnarray*}
W&=&\sum _{j,j_z,\xi}[j(j+1)-j_z^2] \rho _{j,j_z,\xi}\\
%&\leq & \sum_{j,\xi}(j(j+1)) \rho _{j,0,\xi}+\sum _{j,j_z\neq 0,\xi}(j_{\mathrm{M}}(j_{\mathrm{M}}+1)-j_z^2)\rho _{j,j_z,\xi}\\
&\leq & \sum_{j,\xi}j(j+1) \rho _{j,0,\xi} +\sum _{j_z\neq 0} [j_{\mathrm{M}}(j_{\mathrm{M}}+1)-j_z^2] p_{j_z}.
\end{eqnarray*}
The first term on the right-hand side is again bounded as
\begin{eqnarray*}
&&\sum_{j,\xi}[j(j+1)] \rho _{j,0,\xi}\\
%&=& j_{\mathrm{M}}(j_{\mathrm{M}}+1)\rho _{j_{\mathrm{M}},0,1}+\sum _{j\neq j_{\mathrm{M}},\xi}j(j+1)\rho_{j,0,\xi}\\
&\leq & j_{\mathrm{M}}(j_{\mathrm{M}}+1)\rho _{j_{\mathrm{M}},0,1}+(j_{\mathrm{M}}-1)j_{\mathrm{M}}\sum _{j\neq j_{\mathrm{M}},\xi}\rho_{j,0,\xi}\\
&\leq &  j_{\mathrm{M}}(j_{\mathrm{M}}+1)\rho _{j_{\mathrm{M}},0,1}+(j_{\mathrm{M}}-1)j_{\mathrm{M}} (p_0-\rho _{j_{\mathrm{M}},0,1}).
\end{eqnarray*}

Using these inequalities, the fidelity has the following lower bound:
\begin{equation*}
\frac{W}{2j_{\mathrm{M}}}-\frac{j_\mathrm{M}-1}{2}p_0-\sum _{j_z\neq 0} \left( \frac{j_\mathrm{M}+1}{2}-\frac{j_z^2}{2j_{\mathrm{M}}}\right) p_{j_z} \leq\mathrm{F}.
\end{equation*}

\end{document}